

\def\bi{\bigskip}
\def\sm{\smallskip}

\def\ct {\centerline}
\def\noi{\noindent}
\def\ii{\'{\i}}

\def\sm{\smallskip}


\magnification 1200
\baselineskip 15pt
\null
\vskip 1truecm
\ct {\bf FROM A RELATIVISTIC POINT PARTICLE}
\ct {\bf TO STRING THEORY}
\bi
\ct {\it by}
\bi
\ct{\it J.A. Nieto {\footnote {*}{\sevenrm Electronic address
nie@zeus.ccu.umich.mx}}}
\ct{\it Escuela de Ciencias F\ii sico-Matem\'aticas de la}
\ct{\it Universidad Michoacana, Apartado Postal 749,}
\ct{\it {58000, Morelia Michoac\'an, M\'exico.}}
\ct{\it and}
\ct{\it Instituto de F\ii sica de la Universidad de Guanajuato}
\ct{\it Apartado Postal E-143, 37150,}
\ct{\it Le\'on Guanajuato, M\'exico}
\vglue 1.5cm
\noi PACS: 12.10.6q
\bi\bi
\ct{\bf Abstract}
Using a classical action associated to a point-particle in (1+1)-dimensions
the classical string theory is derived. In connection with this result two
aspects are clarified: First, the point particle in (1+1)-dimensions is not an
ordinary relativistic system, but rather a some kind of a relativistic top;
and second, through the quatization of such a kind of top the ordinary string
theory is not obtained, but rather a $\sigma$-model associated to a
non-compact group which may be understood as an extended string theory.
\vglue 4.2cm
August 1995\par
\vfill\eject
The central idea of this work is to rise the question whether the classical
string theory [1] may be derived as a first quantization of a point particle
system. At first sight this idea may appear rather strange, since it is known
that the concept of a string was proposed precisely as one of the simplest
extention of the concept of a relativistic point particle.
In this sence I should mention that what I have in mind is not an ordinary
relativistic point particle which it is extended presumibly by the concept
of a string but rather a special kind of a point particle moving not in an
Minkowsky space-time of (3+1)-dimensions but in a non-ordinary space-time
of (1+1)-dimensions.
\bi
In fact, I show in this article that by quantizing a special kind of a
point particle, which in an special gauge turns out to be the analog of a
relativistic top [2] in (1+1)-dimensions, an extended version of the classical
string theory is derived. In order to distinguish the point particle refering
in this article and the ordinary relativistic point particle it becomes very
convenient to put a name to the new physical system. Further, in order to
emphasize the fact that it is a special kind of a point particle system
"living" in a some kind of dual world of one "space" and one "time", that
is in a non-ordinary ``space-time" of (1+1)-dimensions, I called
``quatl" (From n\'ahuatl word c\'oatl meaning serpent, twin or dual [3]).
\bi
Let me try to clarify still more the main idea of this work. For this purpose
assume that the motion of a Newtonian point particle in an Euclidean space of
3-dimensions, a relativistic point particle in a Minkowsky space-time of
(3+1)-dimensions, and a string in a Minkowsky space-time of (d+1)-dimensions
is described by the position coordinates $x^i (t), x^\mu (\tau)$ and $x^{\hat
\mu}  (\tau, \sigma)$
respectively, where $\tau$ is an arbitrary time-like parameter and $\sigma$ is
an arbitrary parameter used to denote points along the string (here $i =
1,2,3, ~\mu = 0,1,2,3,$ and $\hat \mu = 0,1,2, \ldots, d)$.
The evolution of these coordinates may be expressed symbollically as
\sm
$$x^i (t) \to x^\mu (\xi^0) \to x^{\hat \mu} (\xi^a), \eqno (1)$$
\bi
\noi
where $\xi^0 \equiv \tau$ and $\xi^a = (\tau, \sigma)$, with $a=0,1$. Of
course, this picture will be reflected later on the associated actions, but
even at the level of coordinates one notice important changes.
\bi
In particular one notice that in the transition
$$x^\mu (\xi^0) \to x^{\hat \mu} (\xi^a), \eqno (2)$$
\bi
\noi
not only the indices change $\mu \to {\hat \mu}$, but also the parameters
$$\xi^0 \to \xi^a. \eqno (3)$$
\bi
\noi Looking things in this way one wonders if the change in the parameters
(3) is what it is really important inducing a change in the coordinates (2).
What I would like to mention here is that the simplicity of the transition (2)
may be used to say that strings are one of the simplest extensions of a point
particle.
\bi
Let me make a pause and turn now to the question of the relation between the
coordinates $x^\mu$ and, let say a number of scalar fields $\varphi^i(x^\mu)$.
Simbolically what one would like to analize is the transition
\sm
$$x^\mu \to \varphi^i (x^\mu). \eqno (4)$$
\bi
\noi
One way to understand this transition is to start first with the
classical theory of the relativistic point particle and then quantize the
system using Dirac's constraint Hamiltonian procedure [4]. In this way the
$\varphi^i (x^\mu)$ fields may be interpreted as quantum states.
\bi
I am, now, in  position to clarify more the central idea of this work. First,
notice that just like $\varphi^i (x^\mu)$ are $n$ scalar fields in
(3+1)-dimensions the string coordinates $x^{\hat \mu} (\xi^a)$ may be
understood as (d+1) scalar fields in (1+1)-dimensions. So, just like the
transition (4) one may considers also the symbolic transition
\sm
$$\xi^a \to x^{\hat \mu} (\xi^a). \eqno (5)$$
\bi
\noi
Following the analogy one needs then to develop a classical theory for
a point particle where the motion is described by the coordinates $\xi^a$. In
this way after quantize the system whose motion is described by the
coordinates $\xi^a$ one may interpret the scalar fields
$x^{\hat \mu} (\xi^a)$
as quantum  states.
\bi
In this article  I found that the transition (5) is, in fact, not possible.
This fact force us to look for an alternative prescription. After some
attempts I finally found a transition which mantain more or less the same
ingredients than (5). Symbolically, what I am proposing here, it is to
consider the transition
\sm
$$\xi^a_m \to x^{\hat \mu}_{(\hat \alpha)}(\xi^a_m), \eqno (6)$$
\bi
\noi
with $m=0,1$ and ${(\hat \alpha)}= 0,1,..d$. The coordinates $\xi^a_m$
describe
some kind of a relativistic top in (1+1)-dimensions and are the variables
used to study the motion of what I call quatl. While the quantum states
$x^{\hat \mu}_{(\hat \alpha)}$, arising as a first quantization of the quatl,
admits an interpretation of an element of a non-compact group $SO(d,1)$.
\bi
I need to make some interesting final remarks. According to the transition
(6) the evolution of the parameter (3) may be written now as
\sm
$$\xi^0 \to \xi^a \to \xi^a_m, \eqno (7)$$
\bi
\noi and the evolution of the coordinates (1) may be written now as
$$x^i (t) \to x^\mu (\xi^0) \to x^{\hat \mu} (\xi^a) \to x^{\hat \mu}_{(\hat
\alpha)} (\xi^a_m), \eqno (8)$$
\bi
\noi
Moreover, notice that the expressions (7) and (8) are not of the kind of
evolution that one finds in the case of {\it p}-branes where instead of (7)
and (8) one has
$$\xi^0 \to \xi^a \to \xi^{\hat a}, \eqno (9)$$
\sm
\noi and
$$x^i (t) \to x^\mu (\xi^a) \to x^{\hat \mu} (\xi^a) \to x^{\hat \mu}
(\xi^{\hat a}), \eqno (10)$$
\sm
\noi
with ${\hat a} = 0, 1, 2,..,p$.
\bi
Let me start introducing the string action [1]
\sm
$$S= {1\over2} \int d^2 \xi \sqrt{-g} ~g^{ab} (\xi^c) {\partial x^{\hat \mu}
\over\partial \xi^a} ~{\partial x^{\hat \nu} \over \partial \xi^b} \eta_{{\hat
\mu} {\hat \nu}}, \eqno (11)$$
\bi
\noi where $g_{ab} (\xi^c)$ is a metric on the world-surface swept out by the
string, $\eta_{\hat \mu \hat \nu} =$ diag (-1,...,1) is the Minkowsky
metric in (d+1)-dimensions, and $g$ is the determinant of $g_{ab}$.
\bi
The field equations obtained from (11) are
\sm
$${1 \over {\sqrt - g}} \bigg( {\partial \over \partial \xi^a} \sqrt{- g}
{}~g^{ab} {\partial \over \partial \xi^b} \bigg) x^{\hat \mu} = 0, \eqno (12)$$
\sm
\noi and
\sm
$${\partial x^{\hat \mu} \over \partial \xi^a} {\partial x^{\hat \nu} \over
\partial \xi^b} \eta_{\hat \mu \hat \nu} - {1\over 2} g_{ab} \bigg( g^{cd}
{\partial x^{\hat \mu} \over \partial \xi^c} {\partial x_{\hat \mu} \over
\partial \xi^d} \bigg) = 0, \eqno (13)$$
\bi
\noi which in the gauge $g_{ab} = \eta_{ab} = diag ~(-1,1)$ reduce to
\bi
$$ \eta^{ab} \partial_a \partial_b x^{\hat \mu} = 0, \eqno (14)$$
\sm
\noi and
\sm
$$\partial_a x^{\hat \mu} \partial_b x^{\hat \nu} \eta_{\hat \mu \hat \nu} -
{\eta_{ab}\over2} \bigg( \eta^{cd} \partial_c x^{\hat \mu} \partial_d
x_{\hat \mu} \bigg) = 0. \eqno (15)$$
\bi
Let me now change the subject and turn to explain in more detail the
transition (4).
\bi
The action associated to a relativistic point particle of mass $m$ is given
by
$$S= - m \int d \xi^0 \bigg(- \dot x^\mu \dot x^\nu \eta_{\mu \nu}
\bigg )^{1/2}, \eqno (16)$$
\bi
\noi where $\eta_{\mu \nu} = diag (-1, 1, 1, 1)$ is now the Minkowsky
metric in (3+1)-dimensions and the dot means derivative with respect to
$\xi^0$.
\bi
Defining the linear momentum $P_\mu$ as $P_\mu = {\partial {\cal L} \over
\partial \dot x^\mu}$, where ${\cal L} = - m (- \dot x^\mu \dot x_\mu)^{1/2}$
one  finds the first class constraint
$$P^\mu P_\mu + m^2 = 0. \eqno (17)$$
\sm
\noi Now, using Dirac constraint Hamiltonian procedure one can quantize this
system by promoting the linear momento $P_\mu$ to an operator $\hat P_\mu$
and applying the constraint in allowed physical states $\vert \varphi^i>$. One
gets the field equation
\sm
$$\bigg(\hat P^\mu \hat P_\mu + m^2 \bigg) \vert \varphi^i> = 0,
\eqno (18)$$
\sm
\noi
which in a coordinate basis reduces to
\bi
$$\bigg(\partial^\mu \partial_\mu-m^2\bigg) \varphi^i (x^\mu)=0,
\eqno (19)$$
\bi
\noi
since $\hat P_\mu = - i \partial_\mu$ and $\varphi^i (x^\mu)=
<x^\mu \vert \varphi^i >$.
\sm
Let me now compare (14) and (19). Except for the term $m^2$ these two
expesions have the same structure. So, since (19) is obtained from (18)
using special kind of coordinate basis one could expect to have the
analog of (18) for the case of the string, namely
\sm
$$\hat P^a \hat P_a | x^{\hat \mu} > = 0. \eqno (20)$$
\bi
\noi
At the same time this expresion suggests to write the analog of (17) as
\sm
$$P^a P_a = 0. \eqno (21)$$
\bi
\noi Going backwards, this constraint is expected to be obtained from an
action similar to (16), but now corresponding to a massless point particle
moving in (1+1)-dimensions.
\bi
This picture however does not work because of the second constraint (15).
In fact, since
\sm
$$\hat P^a | x^{\hat \mu}  > = P^a | x^{\hat \mu} >, \eqno (22)$$
\bi
\noi
the constrain (15)  will lead to
$$P_a P_b = 0, \eqno (23)$$
\bi
\noi
and consecuently $P_a$ vanishes.
\bi
Analysing why one obtains (23), one finds that the main reason is
because one has (21). So, one can attempt to introduce a ``mass" term
$\mu^2$ in order to have the constraint
\sm
$$P^a P_a + \mu^2 = 0, \eqno (24)$$
\bi
\noi
instead of (21). But one can check that one still has the problem. In fact,
using (24) the constraint (15) leads to
\sm
$$ P_a P_b +  {1\over 2} \eta_{ab} \mu^2 = 0. \eqno (25)$$
\bi
\noi
But this expression implies again $P_a =0$. The situation can be changed
drastically if one assume that instead of the linear momentum $P^a$ one has
the linear momentum $P^a_m$ where the indice $m$ is some kind of
hidden indice runing from 0 to 1. Considering this change (24) and
(25) become now
\sm
$$P_a^m P_m^a + \mu^2 = 0, \eqno (26)$$
\noi and
$$P_a^m P_{bm} + {1\over 2} \eta_{ab} \mu^2 = 0, \eqno (27)$$
\bi
\noi
where we assume that there is a Matrix $\eta_{mn} = diag ~(-1, 1)$ such than
$P^m_a = \eta^{mn} P_{an}$.
\sm
\noi Note that it is still convenient to have $\mu^2 \neq 0$.
\bi
Altough, one has solved the problem at the level of constraints, one needs,
now, to confront the problem that the quantization of (26) leads to the
quantum equation
$$\bigg( \hat P^m_a \hat P^a_m+\mu^2 \bigg) | x^{\hat \mu} > =0, \eqno (28)$$
\sm
\noi
which differs from the expected quantum string equation (20). The indice $m$
in the momentum $\hat P_a^m$ in (28) is not really the problem since one can
think that it is a hidden indice in (20). What seems to produce now the
problem is the ``mass" term $\mu^2$. This mass term seems to be necessary if
one desires to make consistent the constraint (27), but it spoils the nice
structure of (20). So, what to do. Here I would like to propose an idea which
seems to solve the conflict.
\bi
Assume that instead of the state $| x^{\hat \mu}>$ one has now the state $|
x^{\hat \mu}_{(\hat \alpha)} >$ where $(\hat \alpha)$ is a new indice runing
from $0$ to $d$. This new state deserves a name in order to be distinguished
not only from the state $| x^{\hat \mu} >$ but also from ordinary states $|
\psi >$ in quantum mechanics. According to Dirac's terminology a quantity $|
\psi >$ is called a ket and a state $< \psi |$ is called a bra. I think that
this ket $| \psi >$ is not the same that the ket $| x^{\hat \mu} >$. So I
suggest to make a distinction with these two different kets. In particular I
prefered to call the new state $| x^{\hat \mu}_{(\hat \alpha)} >$ ketzal
(from n\'ahuatl word quetzalli meaning preciouse or treasure [3]).
Thus, using the
ketzal $| x^{\hat \mu}_{(\hat \alpha)} >$ the quantum equation (28)
becomes
$$\bigg( \hat P_a^m \hat P^a_m + \mu^2 \bigg) |
x^{\hat \mu}_{(\hat \alpha)} > = 0, \eqno (29)$$
\sm
\noi
which in coordinates basis may be written as
\sm
$$\bigg( \hat P_a^m \hat P^a_m + \mu^2 \bigg) x^{\hat \mu}_{(\hat \alpha)}
( \xi^a_m ) = 0, \eqno (30)$$
\sm
\noi
with $\hat P^m_a = - i {\partial \over \partial \xi^a_m}$. While the
constraint (27) leads to
\sm
$$\hat P^m_a x^{\hat \mu}_{(\hat \alpha)} \hat P_{bm} x^{\hat \nu}_{{\hat
(\beta)}} \eta_{(\hat \mu \hat \nu)} + {\eta_{ab} \over 2} \eta_{(\hat \alpha
\hat \beta)} \mu^2 = 0, \eqno (31)$$
\bi
\noi
provided one assumes $\hat P^m_a x ^{\hat \mu}_{(\hat \alpha)} =
P^m_a x^{\hat \mu}_{(\hat \alpha)}$ and one imposes the orthonormal
condition
\sm
$$x^{\hat \mu}_{(\hat \alpha)} x^{\hat \nu}_{(\hat \beta)}
\eta_{\hat \mu \hat \nu} = \eta_{(\hat \alpha \hat \beta)}. \eqno (32)$$
\bi
What it is important is that due to (31) and (32) the mass term $\mu^2$
becomes
\sm
$$\mu^2 = -{1 \over d + 1} \bigg( \hat P^m_a \hat x^{(\hat \alpha)}_{\hat \mu}
\hat P^a_m x^{\hat \mu}_{(\hat \alpha)} \bigg). \eqno (33)$$
\sm
\noi
So, (30) and (31) may be written now as
\sm
$$\bigg( \hat P^m_a \hat P^a_m  - {1 \over (d + 1)} \bigg( \hat P^m_a x^{\hat
\nu}_{(\hat \sigma)} \hat P^a_m x^{(\hat \sigma)}_{\hat \nu}\bigg) \bigg)
x^{\hat \mu}_{(\hat \alpha)} (\xi^a_m) = 0, \eqno (34)$$
\sm
\noi and
\sm
$$\hat P^m_a x^{\hat \mu}_{(\hat \alpha)} \hat P_{bm} x^{\hat \nu}_{(\hat
\beta)} \eta_{\hat \mu \hat \nu} - {1 \over 2 (d+1)} \eta_{ab} \eta_{(\hat
\alpha \hat \beta)} \bigg( \eta^{cd} \hat P_c^m x^{\hat \mu}_{(\hat \sigma)}
\hat P_{dm} x^{(\hat \sigma)}_{\hat \mu} \bigg) =0. \eqno (35)$$
\sm
\noi respectively. Using (35) one learns that (34) may also be written as
\sm
$$\hat P^m_a \hat P^a_m x^{\hat \mu}_{(\hat \alpha)} - \hat P^m_a
x^{\hat \lambda}_{(\hat \alpha)} \hat P_{bm} x^{\hat \nu}_{(\hat \beta)}
\eta^{ab} \eta_{\hat \lambda \hat \nu}
x^{\hat \mu (\hat \beta)} = 0. \eqno (36)$$
\sm
\noi Further, multipling (35) by $\eta^{(\hat \alpha \hat \beta)}$ one gets
\sm
$$\hat P^m_a x^{\hat \mu}_{(\hat \alpha)} \hat P_{bm} x^{\hat \nu}_{(\beta)}
\eta^{(\hat \alpha \hat \beta)}-{1 \over 2} \eta_{ab} \bigg( n^{cd}
\hat P^m_c
x^{\hat \mu}_{(\hat \alpha)} P_{dm} x^{(\hat \alpha)}_{\hat \mu} \bigg) = 0.
\eqno (37)$$
\sm
\noi
Finally, since $\hat P^m_a = - i {\partial \over \partial \xi^a_m}$ the
expressions (36) and (37) become
\sm
$$\eta^{ab} \eta_{mn} {\partial^2x^{\hat \mu}_{(\hat \alpha)} \over \partial
\xi^a_m \partial \xi^b_n} - {\partial x^{\hat \lambda}_{(\hat \alpha)} \over
\partial \xi^a_m} ~{\partial x^{\hat \nu}_{(\hat \beta)} \over \partial
\xi^{bm}} \eta^{ab} \eta_{\hat \lambda \hat \nu} x^{\hat \mu (\hat \beta)} =
0, \eqno (38)$$
\noi and
\bi
$${\partial x^{\hat \mu}_{(\alpha)} \over \partial \xi^a_m}
{}~{\partial x^{(\hat \alpha)}_{\hat \mu} \over \partial \xi^{bm}} -
{1\over 2} \eta_{ab} \bigg( \eta^{cd} {\partial x^{\hat \mu}_{(\hat \alpha)}
\over \partial \xi^c_m}~{\partial x^{(\hat \alpha)}_{\hat \mu}
\over \partial \xi^{dm}} \bigg) = 0
\eqno (39)$$
\sm
\noi
\noi respectively. What it becomes very interesting is that (38) and (39) may
obtained from the action
\sm
$${\cal S } = {1\over 2} \int d \xi \sqrt{-g} ~g^{ab} \eta_{mn} {\partial
x^{\hat \mu}_{(\hat \alpha)} \over \partial \xi^a_m} {\partial x^{\hat
\nu}_{(\hat \beta)} \over \partial \xi^b_n}  \eta^{(\hat \alpha \hat \beta)}
\eta_{\hat \mu \hat \nu}, \eqno (40)$$
\bi
\noi
which is an extention of the string action (11) and admits an interpretation
of a $\sigma$-model associated to the non-compact group SO(d,1). In
this interpretation the coordinates $x^{\hat \mu}_{(\hat \alpha)}$ are
elements of SO(d,1). According to this development, the mass term $\mu^2$
becomes only a mathemathical artifice.
\bi
One needs still to ask ourself from which action the constraints (26) and
(27) associated to the quatl may be obtained. Consider the action
\sm
$$S= \int d \eta ~\dot \xi^a_m P^m_a + {\lambda \over 2} \bigg(\sqrt{-g}
{}~g^{ab} P^m_a P_{bm} + \mu^2\bigg), \eqno (41)$$
\sm
\noi
where $\eta$ is an arbitrary parameter and $\lambda$ is a lagrange multiplier.
Varing this action with respect to
$\lambda$, one obtains
$$\sqrt{-g} g^{ab} P^m_a P_{bm} + \mu^2 = 0, \eqno (42)$$
\noi
and varing with respect to $g^{ab}$ one obtains
$$P_a^m P_{bm} - {1 \over 2} g_{ab} (g^{cd} P_c^m P_{dm}) = 0. \eqno (43)$$
The constraints (42) and (43) reduces to (26) and (27) respectively when one
fixes the gauge $g_{ab} = \eta_{ab}$.
\sm
The corresponding action in second order form will be
\sm
$$S = {1\over 2} \int d \eta \bigg( - \sqrt{-g} g^{ab} \dot \xi_{am}
\dot \xi^m_b \bigg)^{1/2}, \eqno (44)$$
\sm
\noi
Note that this action is Weyl invariant $(g_{ab} \to \Lambda g_{ab})$ and
it is invariant under re\-pa\-ra\-me\-tri\-za\-tion  $\xi^c_m \to \xi^{'c}_m =
\xi^{'c}_{m} (\xi^a_n)$.
\bi
Finally, one can attempt to give a physical meaning to the coordinates
$\xi^a_m$ fixing the gauge. Assume the parameter $\xi^a_m$ satisfy the
constraint
\sm
$$\xi^a_m \xi_{an} = \eta_{mn}. \eqno (45)$$
Then the quatl admits an interpretation of a special kind of a top. In fact
let us introduce the angular momentum,
$$\Sigma^{ab} = \xi^a_m P^{mb} - \xi^b_m P^{ma}. \eqno (46)$$
\sm
\noi
Using (45) one finds $P^a_m = \Sigma^{ab} \xi_{bm}$ and therefore the
constraints (42) and (43) become
$$\bigg( \Sigma^{ab} \Sigma_{ab} + \mu^2 \bigg) = 0, \eqno (47)$$
$$\Sigma_{ac} \Sigma^c_{~b} - {1\over 2} g_{ab} \bigg(
\Sigma^{cd} \Sigma_{cd}
\bigg) = 0. \eqno (48)$$
\sm
But, now, by definition $\Sigma^{ab}$ is antisymmetric and
therefore in (1+1) dimensions should be proportial to the Levi-Civitta
tensor in two dimensions $\epsilon^{ab}$, that is,
\sm
$$\Sigma^{ab} = \Sigma \epsilon ^{ab}, \eqno (49)$$
\sm
\noi
where $\Sigma$ is an arbitry factor.
\sm
\noi Sustituting (49) into (48) one gets
\sm
$$\epsilon_{ac} \epsilon^c_b - g_{ab} = 0. \eqno (50)$$
\sm
\noi
But $\epsilon_{ac} \epsilon^c_b = \eta_{ab}$, so in the gauge (45)
$g_{ab}$ reduces to $\eta_{ab}$, and therefore (45) corresponds to the
orthonormal gauge $g_{ab} = \eta_{ab}$ and (48) becomes an identity. A
constraint
of the form  (47) and (48) may be obtained from a first order lagrangian
$${\cal L} = - \sigma^{ab} \Sigma_{ab} + {\lambda \over 2} \bigg( \sqrt{-g}
g^{ab} \Sigma_{ac} \Sigma^c_{~b} + \mu^2 \bigg), \eqno (51)$$
\noi where
$$\sigma^{ab} = \xi^a_{~m} \dot \xi^{bm}, \eqno(52)$$
is the angular velocity. This lagrangian admits an interpretation of some kind
of relativistic top [2].
\bi
Let me summarize the results and make some final remarks.
\bi
Starting with an action associated to special kind of point particle called
quatl I have derived a generalized $\sigma$-model associated to the
non-compact group SO(d,1), which may be understood as an extension of the
string theory. Looking string theory as an extension of this $\sigma$-model
one arrives to the conclusion that the usual first quantization procedure in
the string theory may be interpreted as a second quantization, and therefore,
quantum string field theory should correspond to a third quantization. This
results agree with the suggestion [5] that "classical string theories do not
exist".
\bi
I need to make just a final comment in connection with the new names suggested
in this article. First, since I call the states
$|x^{\hat \mu}_{(\hat \alpha)}>$ ketzal and these states apply to the
constraints corresponding to what I call quatl, then, the complete theory
developed in this article should be called Ketzalquatl or Quetzatc\'oatl (from
n\'ahuatl meaning plumed serpent or precious Twin [3]). Of course,
I should be respectfull about this name since is one of the main deities in
Mesoamerican cultures. But I think that the theory presented here instead
to reduce reenforces the importance of this deity.
\bi
I am grateful to Claudia Rodr\ii guez for editing this
manuscript in tex.
\bi\bi
\ct {\bf REFERENCES}
\bi\bi\bi
\item {[1]} For a Review of string Treory and an extensive list of references
	   see M. Green, V. Schwartz and E. Witten: Superstring Theory, Vol. I
	   and II (Cambridge University Press, 1987); M. Kaku: Introduction
	   to Superstring (Springer-Verlag, New York, N.Y., 1990).
\bi
\item {[2]} A.J. Hanson and T. Regge: Ann. Phys. (N.Y.), 87, 498 (1974)
\bi
\item {[3]} R. Sim\'eon: Diccionario de la lengua N\'ahuatl o Mexicana (Siglo
	     veintiuno editores, colecci\'on am\'erica nuestra, 1988).
\bi
\item {[4]} M. Henneaux and C. Teitelboim, ``Quantization of Gauge Systems",
	    Princenton Univ. Press, New Jersey, 1992.
\bi
\item {[5]} J.M. L\'opez, M.A. Rodr\ii guez, M. Socolovsky, and J.L.
	    V\'azquez, ``Do classical strings exist?" preprint CINVESTAV
	    1988.

\end